\documentclass[useAMS,usegraphicx]{mn2e}
\usepackage{times}
\newif\ifAMStwofonts
\AMStwofontstrue

\def\ngc{{NGC 1068}}
\def\mkn{{Mkn 3}}

\def\le{{L_{\rm Edd}}}

\def\xmm{{\it XMM-Newton}}
\def\chandra{{\it Chandra}}

\def\et{{et al.\ }}

\def\asca{{\it ASCA}}
\def\sax{{\it BeppoSAX}}

\newcommand{\ls}{\mathrel{\hbox{\rlap{\hbox{\lower4pt\hbox{$\sim$}}}\hbox{$<$}}}}
\newcommand{\gs}{\mathrel{\hbox{\rlap{\hbox{\lower4pt\hbox{$\sim$}}}\hbox{$>$}}}}

\def\arcs{{\hbox{$^{\prime\prime}$}}}

\def\H0{{\rm ~km~s^{-1}~Mpc^{-1}}}

\def\et{{et al.}}

\def\deg{^\circ}

\title[X-ray spectrum of Seyfert 2]
        {X-ray reflection in the nearby Seyfert 2 galaxy \ngc}
\author[K.A.Pounds \et]
        {Ken Pounds
	and Simon Vaughan \\
	Department of Physics and Astronomy, University of Leicester,
Leicester, LE1 7RH, UK\\}

\date{Accepted: 6/2/2006; Submitted: 20/01/2006, in original form 06/12/2005}
\pagerange{\pageref{firstpage}--\pageref{lastpage}}
\pubyear{2006}
\begin{document}
\maketitle
\label{firstpage}

\begin{abstract}  
We use the full broad-band \xmm\ EPIC data to examine the X-ray spectrum of the nearby Seyfert 2 galaxy \ngc, previously shown
to  be complex with the X-ray continuum being a sum of components reflected/scattered from cold (neutral) and warm (ionised)
matter, together with associated emission line spectra. We quantify the neutral and ionised reflectors in terms of the
luminosity of the hidden nucleus. Both are relatively weak, a result we interpret on the  Unified Seyfert Model by a near
side-on view to the putative torus, reducing the visibility of the illuminated inner surface of the torus (the cold reflector),
and part of the ionised outflow. A high inclination in \ngc\ also provides a natural explanation for the large (Compton-thick)
absorbing column in the line-of-sight to the nucleus. The emission line fluxes are consistent with the strength of the neutral
and ionised continuum components, supporting the robustness of the spectral model. 

\end{abstract}


\section{Introduction}

At X-ray energies the low redshift Seyfert 2 galaxies \ngc\ and \mkn\ exhibit very `hard' broad-band spectra due to strong
absorption  of the intrinsic X-ray continuum by a large column density of intervening cold matter (the torus?). Previous analyses
of \xmm\ EPIC spectra (\ngc: Matt \et\ 2004, hereafter M04; \mkn: Pounds \et\ 2005, hereafter P05) have confirmed earlier
suggestions that the observed X-flux over the 3-10 keV band in these two nearby AGN is dominated by indirect radiation
`reflected' into the line  of sight. We use the generic term `reflection' here to describe the re-direction of the hidden nuclear
X-ray continuum into the line  of sight by electron scattering from neutral or near-neutral (`cold') matter, and from ionised
(`warm') gas. Line emission arises from the same cold and warm reflectors as fluorescent (inner shell) transitions and by
recombination/radiative decays, respectively.  High resolution grating spectra from both \xmm\ and \chandra\ have confirmed the
soft X-ray band in both \ngc\ and \mkn\ to be dominated by line emission from a photoionised/photoexcited gas (\ngc: Kinkhabwala
\et\ 2002, Ogle \et\ 2003; \mkn: Sako \et\ 2000, P05), with observed outflow velocities of 400--500 km s$^{-1}$. 

Depression of the strong nuclear continuum in Seyfert 2 AGN makes them particularly suitable for studying the circumnuclear
material, by both X-ray absorption and emission spectra. The structure and dynamics of the putative torus can be probed in terms
of absorption and `cold reflection', including fluorescence line emission, while the `warm reflector' required by both the
polarised broad optical lines and the soft X-ray line emission allows the ionised gas outflow to be quantified.

The aim of this paper is to assess the broad-band \xmm\ EPIC spectrum of \ngc\ in terms of the Unified Model of AGN (Antonucci
1993), and compare the outcome with the similar study of \mkn, which P05 concluded was probably a rather special case, being
viewed `just over the edge' of the torus. 

\section{EPIC observation of \ngc\ and data reduction}

\ngc\ was observed by \xmm\ on 2000 July 29-30. As the MOS1 camera was operated in full frame mode and suffered significant
signal pile-up, we use X-ray spectra solely from the EPIC pn (Str\"{u}der \et\ 2001) and MOS2 (Turner \et\ 2001) cameras.  The pn
camera was operated in the large window mode and MOS2 in the small window mode, both with the medium filter, and pile-up was
found to be unimportant. The X-ray data were screened with the SAS v6.3 software and events corresponding to patterns 0-4 (single
and double pixel events) were selected for the pn data and patterns 0-12 for the MOS data. We separately obtained single pixel pn
data to quantify the (small) energy shift in the double pixel counts (P05), but retained the full data set for the subsequent
analysis as it provides significantly better statistical quality in the important Fe K band and above. EPIC source counts were
taken within a circular region of 40\arcs\ radius about the centroid position of \ngc, with the background being taken from a
similar region, offset from but close to the source. The net exposures available for spectral fitting after removal of some high
background data were 61.9 ks (pn) and 68.7 ks (MOS2).  Since no obvious variability was evident throughout the observation,
spectral data were then integrated over the full exposures and binned to a minimum of 20 counts per bin, to facilitate use of the
$\chi^2$ minimalisation technique in spectral fitting.  Spectral fitting was based on the Xspec package (Arnaud 1996) and  all
spectral fits include absorption due to the \ngc\ line-of-sight Galactic column of $N_{H}=3.5\times10^{20}\rm{cm}^{-2}$ (Dickey
and Lockman 1990). Errors are quoted at the 90\% confidence level ($\Delta \chi^{2}=2.7$ for one interesting parameter).    

\begin{figure}                                                          
\centering                                                              
\includegraphics[width=4.7cm, angle=270]{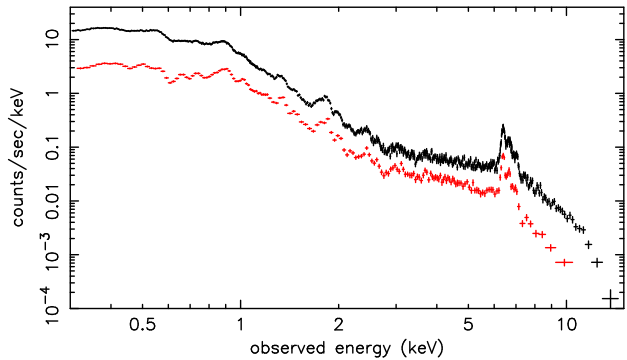}                     
\caption                                                                
{Background-subtracted count spectra from the EPIC observation of \ngc. The pn data are shown in black and those from 
MOS2 in red}      
\end{figure}

\section{The hard (3.5-15 keV) EPIC spectrum} 

The integrated pn and MOS2 count spectra of \ngc\ are reproduced in figure 1 and show considerable structure at both high and low
energies. A simple power law fit to the data above 3 keV finds it to be `hard', with a photon index $\Gamma$$\sim$1. However, an
observation of \ngc\ to 100 keV with \sax\ confirmed the presence of a `normal' Seyfert continuum source  ($\Gamma$$\sim$1.8;
Nandra and Pounds 1994) obscured by a Compton thick absorber (N$_{H}$$\ga10^{26}$cm$^{-2}$; Matt \et\ 1997). Those authors, and
Iwasawa \et (1997) using \asca\ data, interpreted  the X-ray flux observed at $\sim$3--10 keV as arising by reflection/scattering
of the nuclear power law continuum into the line of  sight from cold  matter, perhaps the inner face of the putative torus, with
an additional component from the warm (highly ionised) medium also responsible for the polarised broad optical lines. Here we
follow that analysis (see also M04) by fitting the 3.5-15 keV pn and MOS data, excluding the emission features at $\sim$6--8 keV, with
2 continuum components, one scattered into the line of sight by the warm reflector, retaining the slope of the intrinsic power
law but diluted in accord with the optical depth and covering factor  of the ionised gas, together with a cold reflection
component, modelled in Xspec by PEXRAV (Magdziarz and Zdziarski 1995). This dual reflection model yields an acceptable continuum
fit ($\chi^{2}$=523/490 dof), with photon index $\Gamma$=2.1$\pm$0.1 now within the range for a type 1 Seyfert nucleus
(more specifically of a Narrow Line Seyfert 1; see Discussion), supporting the initial assumption that the ${\it observed}$ hard
spectrum of \ngc\ is reflection-dominated. Applying this continuum model to the whole 3.5-15 keV
data shows the highly significant excess flux at $\sim$6--8 keV (figure 2). 

\begin{figure}                                                          
\centering                                                              
\includegraphics[width=4.7cm, angle=270]{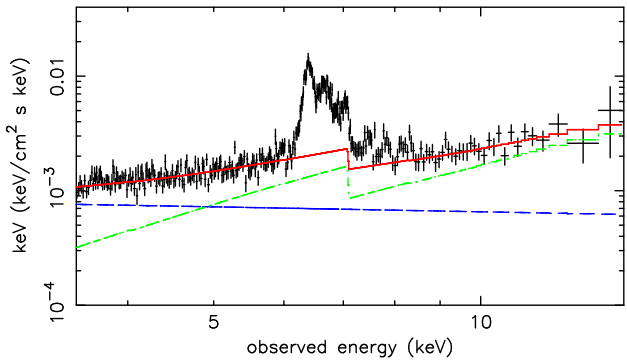}                     
\caption                                                                
{Fit to the EPIC data of \ngc\ over the 3--15 keV spectral band with continuum components from cold (green)
and warm (blue) reflectors shows excess emission at $\sim$6--8 keV. Only the pn data are shown for clarity}      
\end{figure}

We then fit that excess with a sequence of Gaussian emission lines, with line energy and flux as free parameters and line widths
initially tied. The number of lines that can be resolved depends on the EPIC resolution and the actual line widths. M04 found 9
lines in their fit, assuming intrinsically narrow lines. We find a best-fit tied line width $\sigma$=30$\pm$10 eV, with - again -
9 emission lines being statistically significant. Justification of such a complex spectral fit, a tribute to the spectrometric
capability of EPIC, is discussed further in the Appendix.

Our fit includes a single-scattering Compton shoulder ($\sigma$$\sim$80eV) centered at 0.1 keV below the Fe K$\alpha$ line of Fe
(Hatchett and Weaver 1977). There is also marginal evidence for a similar Compton shoulder to Ni K$\alpha$, which we then include
for physical consistency. The addition of those 10 emission lines yields an acceptable overall 3.5--15 keV spectral fit, with 
$\chi^{2}$=858/805 dof.

\begin{figure}                                                          
\centering                                                              
\includegraphics[width=4.7cm, angle=270]{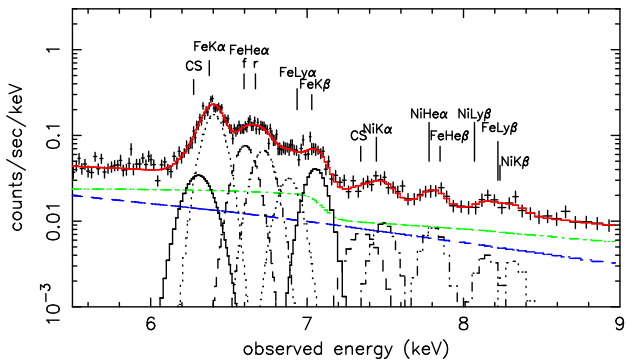}                     
\caption  
{Addition to the continuum fit in figure 2 of emission lines from neutral and ionised Fe and Ni. The zero velocity energy of 
each line, 
in the observer frame, is marked,
showing all the fluorescent lines have at most a small blue shift. The ionised Fe and Ni lines match the rest energies  
less well,
with an indication that several may have substantial velocity shifts}      
\end{figure}

Figure 3 shows the line emission fit in more detail, together with the cold and warm reflection continua. The zero-velocity
energies of the candidate line identifications of  the K-shell emission of Fe and Ni, the only abundant metals with transitions
in this energy band, are also shown in the figure. Fluxes and equivalent widths of the principal lines are included in Table 1.

Fluorescent emission from the cold reflector is clearly seen in the K$\alpha$ lines of Fe and Ni, and K$\beta$ of Fe, with all 3 
fluorescent line energies indicating (at most) a small blue-shift from the neutral line energy. (Re-fitting the line spectrum
using only single pixel pn counts yielded a rest frame energy for Fe K$\alpha$ of 6.415$\pm$0.007 keV.) The relative strength of
the Compton shoulder for the FeK$\alpha$ line, at $\sim$20\%, is consistent with reprocessing in Compton-thick reflecting matter
(Matt 2002). The flux ratio of Fe K$\alpha$ to K$\beta$, of $\sim$4.5, compares with a theoretical ratio of $\sim$6 (Molendi \et\
2003), the strong K$\beta$ line being consistent with the low ionisation state of the `cold' reflector. 

Identifying the line emission from the warm (ionised) reflector is less straightforward, with line blending and an indication of
substantial velocity shifts in some lines. Intriguingly, line shifts have been indicated in previous observations of ionised Fe K
lines in \ngc\ (Iwasawa \et\ 1997, Ogle \et\ 2003), implying velocities of $\sim$2000-5000 km s$^{-1}$. Confirmation of high
velocity/highly ionised gas extending above the putative torus in \ngc\ would be of considerable interest, given growing evidence
for such flows in luminous type 1 AGN (Chartas \et 2002, Pounds \et\ 2003, O'Brien \et\ 2005). We defer further consideration of
this possibility to a subsequent paper (Pounds \et 2006). 

Based on the above  broad band spectral fit we find an ${\it observed}$ 3-15 keV flux for \ngc\ of $6.3\times10^{-12}$ erg
cm$^{-2}$ s$^{-1}$, corresponding (for a red-shift of 0.00379; Huchra \et\ 1999) to a luminosity of $2.2\times 10^{41}$  erg
s$^{-1}$ ($ H_0 = 70 $ km\,s$^{-1}$\,Mpc$^{-1}$). For later reference, over the 2--10 keV band the cold reflection continuum 
luminosity is L$_{cold}$ $\sim$$5.2\times 10^{40}$ erg s$^{-1}$, with L$_{warm}$$\sim$$6.2\times 10^{40}$ erg s$^{-1}$.

\section{Extrapolating the hard EPIC spectral fit of \ngc\ to 0.3 keV}

Extending the 3.5-15 keV spectral fit to lower energies reveals a marked and highly structured `soft excess' (figure 4). The
excess soft X-ray emission is found to have an observed flux (0.3-3 keV) of $1.6\times10^{-11}$~erg cm$^{-2}$ s$^{-1}$. Allowing
for attenuation by the Galactic column of $N_{H}=3.5\times10^{20}$ cm$^{-2}$ increases this by $\sim$10 percent, corresponding to
an intrinsic soft X-ray luminosity over this energy band of $5.6\times 10^{41}$~erg s$^{-1}$. 

\begin{figure}                                                          
\centering                                                              
\includegraphics[width=4.7cm, angle=270]{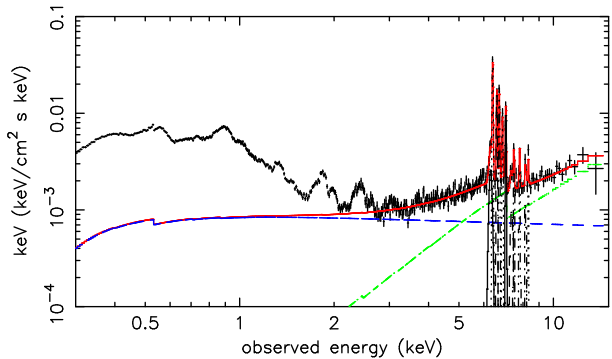}                     
\caption                                                                
{Low energy extension of the 3--15 keV spectral fit, revealing a highly structured soft excess}      
\end{figure}                                                            
 
Similar low energy structure is seen in  both pn and MOS spectra and we choose - as for \mkn\ in P05 - to study the MOS data in
more detail, since in the  soft X-ray band both EPIC cameras have ample statistics while the energy resolution in the MOS CCDs is
better. [The 1$\sigma$ resolution of the MOS at the time of the \ngc\ observation was $\sim$34 eV at 1.5 keV (M.Kirsch,
XMM-SOC-CAL-TN-0018 issue 2.3, 28 July 2004)]. As P05 found in their analysis of \mkn\ the MOS camera is remarkably good in
resolving the main spectral features in a line-dominated spectrum. Although not competing with the RGS spectrum in terms of
energy resolution, the MOS data are a valuable complement to the RGS by virtue of having better count statistics, a sensitivity
extending to higher energies, and potentially providing a measure of line flux in broad wings that would be difficult to distinguish in high
resolution spectra. The quantitative comparison of line and continuum emission provides an important check on the
reflection-dominated spectrum of \ngc.

\begin{figure}                                                          
\centering                                                              
\includegraphics[width=4.7cm, angle=270]{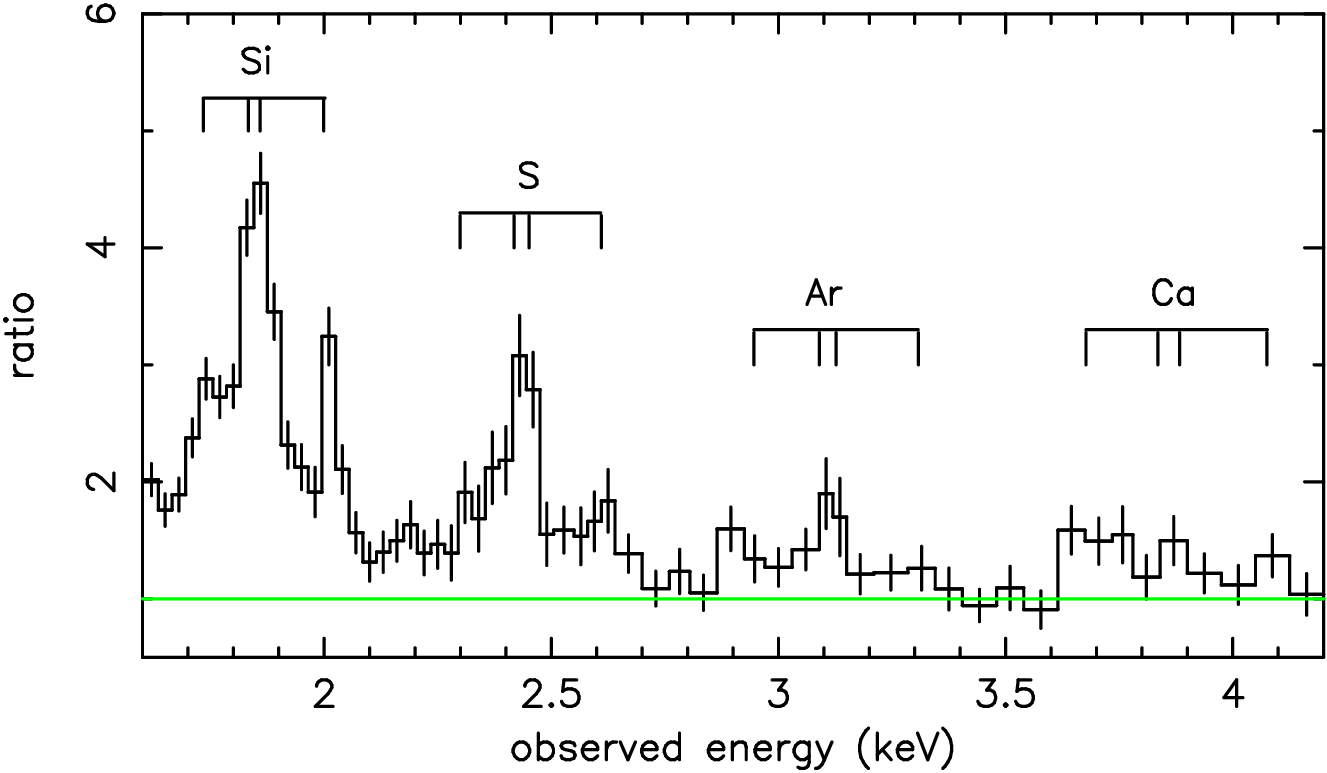}                                                                                                  
\centering                                                              
\includegraphics[width=4.7cm, angle=270]{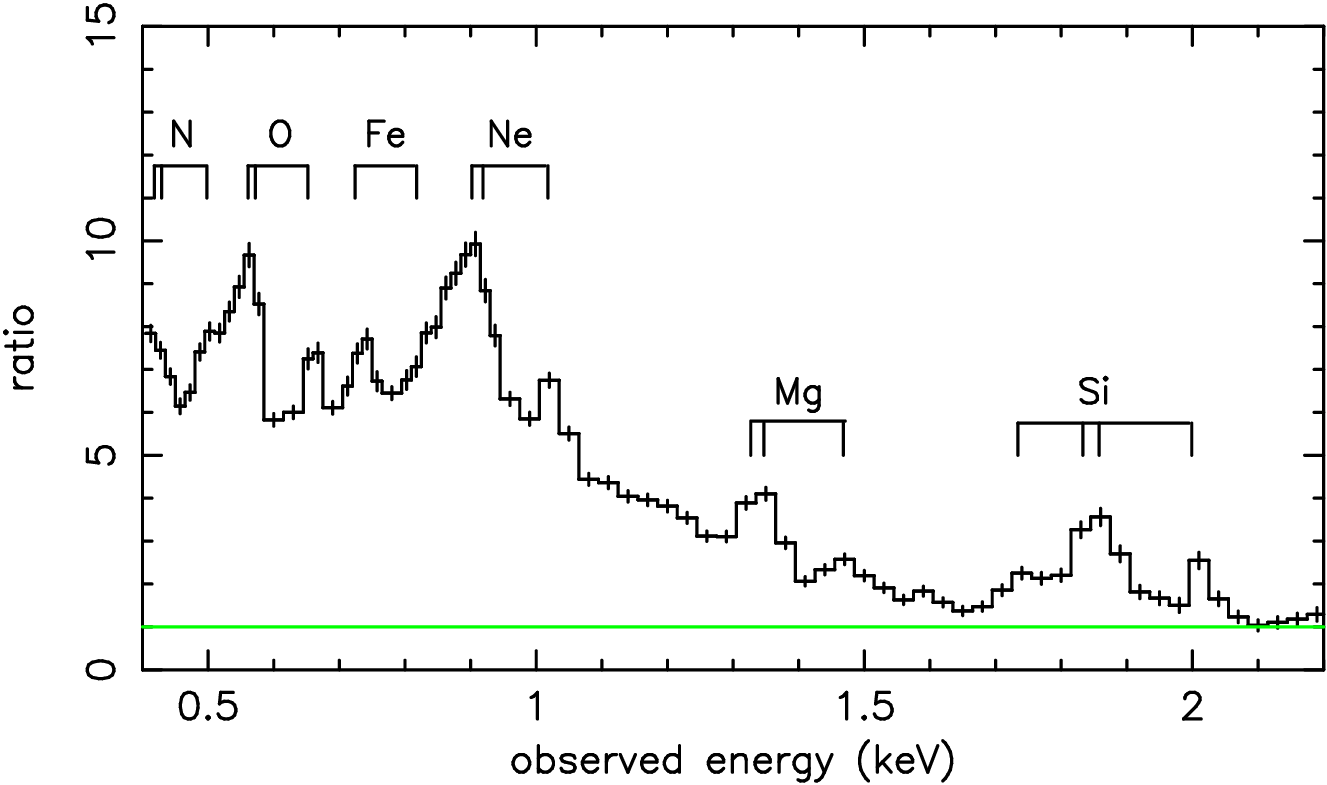}                                         
\caption                                                                
{Identifying features in the EPIC (MOS camera) spectrum of \ngc\ at 1.6-4.2 keV (top) with emission lines of both cold 
and
highly ionised Si, S, Ar and Ca. For each element the markers indicate the rest energies of the K$\alpha$, He-like
1s-2p forbidden and resonance, and Ly$\alpha$ lines respectively. The lower panel shows the
MOS spectrum at 0.4-2.2 keV with the principal emission lines of 
highly ionised N, O, Ne, Mg, and Si, together with fluorescent Si K$\alpha$. The higher resolution RGS data shows
the features identified with Fe are
blends of strong Fe-L lines and the RRC of OVII and VIII}
\end{figure}

Visual examination of the low energy excess in the MOS spectrum of \ngc\ (figure 5) confirms that the neutral K$\alpha$ lines of
Si, S, Ca and Ar are detected, with equivalent widths (compared to the cold reflection continuum), listed in Table 1, broadly
consistent with fluorescence in solar abundance matter (Matt, Fabian and Reynolds 1997). The resonance emission lines of highly
ionised ions of Ca, Ar, S, Si, Mg, Ne and O can also be resolved in the MOS data, though comparison with the higher resolution
RGS data for \ngc\  (Kinkhabwala \et\ 2002) shows many weaker lines, particularly of Fe-L and higher transitions of the He- and
H-like ions, that are blended in the MOS spectra below $\sim$1.5 keV. For example, the peaks identified with Fe-L lines in figure
5 are seen in the RGS data to be blended with equally strong radiative recombination continua (RRC) of OVII and VIII.
Nevertheless, we expect the flux and equivalent widths of the resonance lines of He- and H-like ions of S and Si (also listed in
Table 1) to be reasonably well determined.

We see no significant blue- or red-shifts in the lines resolved
by the MOS, consistent with the higher resolution RGS data, from which Kinkhabwala \et\ find an outflow velocity of
$\sim$400-500 km s$^{-1}$ and line widths of similar magnitude. In particular, there is no evidence in the low and moderate
ionisation gas for the high velocities suggested for the highly ionised Fe K lines. 

A key assumption in deducing the line EWs is that the incident continuum is correctly modelled as in figure 4. For example, a
strong up-turn in the soft X-ray continuum is often assumed in fitting Seyfert 1 spectra (although the physical reality of this
`soft excess' has recently been questioned by Gierlinski and Done 2004). It is instructive in this respect to compare the excess
soft X-ray flux in figure 4 with the sum of the line fluxes resolved by the RGS (Kinkhabwala \et\ 2002). Between 0.35-2 keV
the narrow emission lines resolved by the RGS account for $\sim$70\% of the total soft excess flux over the same energy band 
in figure 4.  Allowance for possible broad wings to the principal lines, and for faint or blended lines not listed in the RGS
analysis could easily provide the difference. That suggests any continuum `soft excess' in in the intrinsic X-ray spectrum of 
\ngc\ is small or absent.

\section{Discussion}

\subsection{Visual comparison of the broad-band X-ray spectra of \ngc\ and \mkn.}

Analysis of the EPIC spectrum of \ngc\ over the 3.5--15 keV energy band shows the X-ray continuum to be composed of two 
components, arising from the hidden nucleus being scattered into the line-of-sight by both cold and warm reflectors. This
spectral modelling is markedly different from that of \mkn\ (P05), where the 3--15 keV spectrum is primarily an addition of the
attenuated intrinsic power law and a strong cold reflection component. This difference is clear in a visual comparison of the raw
EPIC data of \ngc\ and \mkn\ (figure 6). The cold-reflection-dominated continuum of \mkn\ is evident in the extremely hard 
spectrum ($\Gamma$$\sim$ --0.5) below $\sim$6 keV, and in the deep absorption Fe K absorption edge  at 7.1 keV. (We see later
that the strong cold reflection in \mkn\ results in it being as bright as \ngc\ near $\sim$6 keV, despite being 3.5 times more
distant.) Above $\sim$10 keV the much softer observed \mkn\ spectrum is a consequence of the underlying $\Gamma$$\sim$1.8 power
law becoming visible through the marginally Compton-thick absorber. In contrast, while the absorbed power law component enhances
the overall Fe K absorption edge in \mkn, the power law component in \ngc\ reflected from highly ionised matter (acting more like
a perfect mirror) has the effect of diluting the observed 7.1 keV edge and softening the continuum below $\sim$6 keV
($\Gamma$$\sim$1).

In the soft X-ray band, below $\sim$3 keV,  visual examination of the EPIC data shows the soft X-ray spectra of \ngc\ and \mkn\
to be similar in both spectral structure and luminosity, although in \mkn\ the flux falls more steeply at the lowest energies.
While the latter difference is partly explained by the larger Galactic column in the line-of-sight to \mkn, some residual
difference suggests a higher mean ionisation parameter for the outflow visible in \mkn, a view supported by the higher ratio of
Ly$\alpha$ to He$\alpha$ line fluxes in Table 1 for \mkn.

\begin{figure}                                                          
\centering                                                              
\includegraphics[width=4.7cm, angle=270]{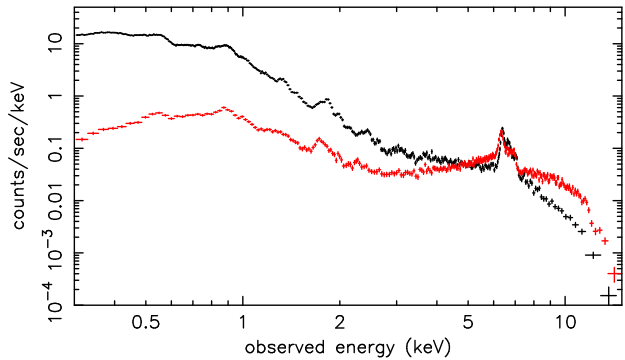}                                                                                                       
\caption 
{Comparison of the raw pn camera data for \ngc\ (black) and \mkn\ (red)}      
\end{figure}

\begin{table*}
\centering
\caption{Comparison of the principal line fluxes and equivalent widths determined from the EPIC data
for \ngc\ and \mkn. Line fluxes are in units of $10^{-6}$~photons 
cm$^{-2}$ s$^{-1}$. The equivalent width (EW) of each observed line is measured against the
corresponding cold or warm reflection component in the continuum model fit. Calculated EWs for the fluorescent lines
are for a solar abundance gas illuminated isotropically, while those for the Fe XXV and FeXXVI resonance lines are
for a column density through the ionised gas of
N$_{H}$= $2\times10^{22}$ cm$^{-2}$. See Sections 5.2 and 5.2 for discussion.}                                                  

\begin{tabular}{@{}lcccccc@{}}
\hline
Line & \ngc\ flux & EW(eV) & \mkn\ flux & EW(eV) & EW$_{calc}$(eV) \\

\hline
Si I K$\alpha$ & 15$\pm$5 & 300$\pm$100 & 10$\pm$2 & 200$\pm$40 & 165 \\
Si XIII (1s-2p)rif & 30$\pm$4 & 140$\pm$20 & 10$\pm$2 & 70$\pm$15 & \\
Si XIV Ly$\alpha$ & 15$\pm$5 & 75$\pm$25 & 8$\pm$2 & 140$\pm$35 & \\
S I K$\alpha$ & 4$\pm$2 & 110$\pm$55 & 7$\pm$3 & 140$\pm$60 & 230 \\
S XV (1s-2p)rif & 12$\pm$4 & 60$\pm$20 & 3$\pm$1.5 & 60$\pm$30 & \\
S XVI Ly$\alpha$ & 6$\pm$3 & 30$\pm$15 & 3$\pm$1.5 & 65$\pm$30 & \\
Fe I K$\alpha$ & 41$\pm$7 & 600$\pm$100 & 30$\pm$3 & 1100$\pm$100 & 1000 \\
Fe I K$\alpha$ CS & 9.5$\pm$ & 130$\pm$70 & 6$\pm$1 & 200$\pm$40 & \\
Fe XXV (1s-2p)r & 17$\pm$3 & 340$\pm$60 & $\leq$2 & $\leq$1400 & 350 \\
Fe XXVI Ly$\alpha$ & 8$\pm$1.5 & 250$\pm$45 & $\leq$2 & $\leq$1400 & 250 \\
Fe I K$\beta$ & 11$\pm$1.5 & 220$\pm$30 & 3$\pm$1.5 & 110$\pm$55 & \\
Ni I K$\alpha$ & 3$\pm$1 & 140$\pm$50 & 2$\pm$1 & 60$\pm$30 & 90 \\

\hline
\end{tabular}
\end{table*}

Table 1 lists the estimated fluxes and equivalent widths (EW) of the the principal emission lines that can be adequately resolved in the
EPIC data for \ngc\ and \mkn. In the Fe and Ni K bands our derived fluxes agree well with those in MO4, although our calculated
equivalent widths are a factor $\sim$2 lower. While the softer X-ray line fluxes have substantial errors, the derived values are in
general agreement with those from the \chandra\ HETG observation reported by Ogle \et\ (2003). For our present purposes, to assess the
reflection-dominated X-ray spectrum of \ngc, the measured line equivalent widths - and their comparison with simple theory - are of
most interest.

Taking first the fluorescent lines of Ni, Fe, S and Si, all have a large EW, within a factor $\sim$2 of the prediction for
reflection from cold, optically thick matter of solar abundance. Taken as a group, the measured values are consistent with the relative
strength of the cold reflector in our continuum fit for \ngc, as was found by P05 for \mkn.

For the warm reflector, the principal emission lines of ionised S and Si are seen with similar EWs in both \ngc\ and \mkn. The 
non-detection of He- and H-like Fe emission in \mkn\ can be understood partly by the
dominance of the cold reflection and intrinsic power law continua (over the ionised reflector) in the Fe K band for that source.  
In the case of \ngc\ those resonance lines are detected, but with EWs much lower than the ($\sim$2--3 keV) values predicted for 
scattering of the hidden X-ray continuum from optically thin gas (Matt \et 1996). Included in the final column of Table 1 are reduced values of the 
Fe XXV and XXVI resonance line EWs assuming a column density through the warm reflector of N$_{H}$= $2\times10^{22}$ cm$^{-2}$, as
derived in the next Section.  

\subsection{Measuring the strength of reflection in \ngc\ and \mkn}

Implicit in the above discussion is that the reflection components have a simple relationship to the irradiating continuum. The cold
reflection will be a product of L$_{X}$$\eta$$\Omega$/4$\pi$, where L$_{X}$ is the intrinsic nuclear X-ray luminosity,  $\eta$
is the mean albedo and $\Omega$/4$\pi$ is the fractional solid angle of illuminated cold matter in our line of sight. Though
the  incident spectral index and reflection geometry are also significant, as is the relative Fe abundance, the principal determinant
of the strength of the reflection, relative to the incident flux, is the solid angle of reflector in view. For the warm
reflector, the observed continuum luminosity will, for an optically thin plasma, be a product of L$_{X}$$\tau$$\Omega$/4$\pi$, 
where $\Omega$/4$\pi$ is
the covering fraction (of the source) by matter of scattering optical depth $\tau$.

Where the intrinsic power law component is detected, as in \mkn, the PEXRAV model in Xspec yields a measure of
$\Omega$/2$\pi$, termed the reflection factor R. P05 found R = 1.7$\pm$0.1, consistent with the strong cold reflection
in \mkn. A similar estimate of R is not possible  for \ngc\ where the intrinsic X-ray source is not seen in the
\xmm\ band. 

Fortunately, the 12$\mu$m flux has been shown to be a good indicator of the nuclear bolometric luminosity of Seyfert galaxies of
both type 1 and type 2 (Spinoglio and Malkan 1989), while the bolometric and 2--10 keV luminosities are also often closely
coupled (e.g. Marconi \et 2004). For \ngc\ the 12$\mu$m flux yields L$_{bol}$$\sim$$6.4\times 10^{44}$ erg  s$^{-1}$, which in
turn allows an estimate of the unseen X-ray luminosity (Marconi \et 2004) of L$_{2-10}$$\sim$$2.3\times  10^{43}$ erg
s$^{-1}$. Comparing that figure with our measured value for the cold reflection continuum  L$_{cold}$ $\sim$$5.2\times 10^{40}$
erg s$^{-1}$, taking $\eta$$\sim$0.02 (George and Fabian 1991), we find $\Omega$/4$\pi$$\sim$0.11. Such a small value might be
seen in a geometry where the irradiated inner surface of the torus is largely hidden by a side-on view angle. In \ngc\ that
conclusion would be consistent with the water maser observations (Gallimore \et\ 1996; Greenhill \et\ 1996) which indicate a
toroidal disc, of $\sim$1 pc scale, at an inclination $\ga$80$\deg$.  

In conceptual terms the extremely strong cold reflection in \mkn\ creates more difficulty, implying that we are seeing
unattenuated  reflection from  matter completely surrounding the nuclear source. One possibility might be that we have
underestimated the intrinsic X-ray flux. To check that, we again use the 12$\mu$m flux to estimate the bolometric luminosity of
\mkn, finding  L$_{bol}$ $\sim$$2\times 10^{44}$ erg s$^{-1}$. Assuming the 25-fold ratio of L$_{bol}$ to L$_{2-10}$ (Marconi
\et\ 2004), we then retrieve the intrinsic luminosity as L$_{2-10}$ $\sim$$8\times 10^{42}$ erg s$^{-1}$, very close to the
value obtained from the \xmm\ analysis of \mkn. Thus it appears that the strong cold reflection in \mkn\ requires a rather
special geometry, where a viewing angle just cutting the near-side edge of the torus, as proposed in P05, sees directly most of
the illuminated far inner wall of the putative torus. In addition, transmission through a column density $N_{H}$$\sim10^{24}$
cm$^{-2}$ of cold matter on the  near side of the torus may provide a significant additional component to both continuum and Fe
K$\alpha$ line, which might also explain the significant blue-shift of the Fe K$\alpha$ line in \mkn\ (PO5).    

The strength of the warm reflected continuum provides a measure of the ionised gas extending above the obscuring torus.  From
the spectral deconvolution of \ngc\ we  had, over the 2--10 keV band, L$_{warm}$$\sim$$6.2\times 10^{40}$ erg s$^{-1}$. Compared
with the incident luminosity of L$_{2-10}$$\sim$$2.3\times 10^{43}$  erg s$^{-1}$, we find $\tau$.C $\sim$0.25 \%. This value is
close to that found from optical spectroscopy (Miller \et\ 1991) and photoionisation modelling (e.g. Storchi-Bergmann \et\ 1992)
of the NLR.  The effect of the higher scattering cross section in resonance lines is seen clearly below $\sim$2 keV where strong
line emission is responsible for the soft X-ray hump seen in figure 4 (Kinkhabwala \et \ 2002). Assuming the continuum
scattering fraction of 0.25\% and a covering factor C$\sim$0.1 (Miller \et\ 1991), we deduce a column density through the ionised gas of
N$_{H}$$\sim$$3\times10^{22}$ cm$^{-2}$, which would be sufficient to strongly depress resonance line fluxes without a significant velocity shear in
the outflow (Matt \et\ 1996). Such a substantial reduction is indeed indicated in the measured equivalent widths of the ionised
reflector resonance lines in Table 1, where we find reasonable agreement with theory (Matt \et\ 1996) assuming a column density of 
N$_{H}$$\sim$$2\times10^{22}$ cm$^{-2}$.

Referring again to figure 6, given that \ngc\ is 3 times more luminous and 3.5 times closer than \mkn, we would expect - with 
identical warm reflectors -  that \mkn\ would  be $\sim$36 times fainter than \ngc\ in the soft X-ray band. Figure 7 shows the
observed difference is closer to a factor $\sim$12 at 1 keV, increasing to $\sim$20 at 0.5 keV, allowing for the larger Galactic
column to \mkn. Thus the line-dominated soft X-ray emissivity in \mkn\ is a factor $\sim$1.8--3 larger than in \ngc. For the
warm continuum component, a comparison with the intrinsic luminosity over the same 2--10 keV band, as above, yields  a
scattering fraction for \mkn\ of 1.1\%. Thus both continuum and line fluxes suggest a factor $\sim$2--4 more ionised matter is
visible in \mkn.

\subsection{An interpretation on the Unified Scheme}

In terms of the unified Seyfert geometry (e.g. Antonucci 1993), where the nuclear source is hidden from direct view at large 
inclinations by an optically thick torus, we now have 3 independent deductions from the X-ray spectral analysis which all
suggest that \ngc\ is observed near side-on (Compton thick absorption, weak cold ${\it and}$ and relatively weak warm
reflection), in contrast to \mkn, which PO5 concluded is viewed at a large angle to the plane of the torus, with a smaller
column to the nucleus, and a much  larger extent of the illuminated inner surface visible. Again, on that standard picture, the
larger ionised reflection seen in \mkn\ could be a result of seeing further down the throat of the torus and correspondingly
more of the outflowing ionised gas. The spectral difference in the soft emission in the 2 cases suggests the gas closer to the
nucleus is also more highly ionised.

We note, in passing, that integrating such an ionised outflow to much smaller radii would contribute a stronger soft X-ray emission
spectrum in type 1 Seyferts, as suggested recently ( Pounds \et\ 2005b and references therein). The relatively high velocities and
strong saturation in resonance line cores expected in that case could put much of that extra emission into broad wings which would be difficult to
see with  current high resolution instrumentation.

\section{Summary}
The \xmm\ observation of \ngc\ highlights the potential to explore the
local circumstellar matter in type 2 AGN, afforded by the strong suppression of the direct nuclear continuum.

The hard X-ray ($\ga$3 keV) spectra of both \ngc\ and \mkn\ have continuum components arising from reflection from cold matter,
perhaps the putative molecular torus. Both also have a strong emission line at $\sim$6.4 keV consistent with fluorescence from
the same cold matter. While the absorbing column density to the nuclear  X-ray source in \mkn\ is only marginally  Compton-thick
(N$_{H}$$\sim$1.2$\times 10^{24}$cm$^{-2}$), allowing the intrinsic power law to be seen $\ga$8 keV, the cold absorber in \ngc\
is totally opaque in the EPIC band. Paradoxically, the Compton-thick source, \ngc, has the less hard overall spectrum and a much
weaker Fe K edge. Spectral deconvolution confirms that this is due to a relatively strong continuum component reflected from
warm matter into to line of sight in \ngc. We note that, if Compton-thick AGN are a dominant component of the
Cosmic X-ray Background ($\Gamma$$\sim$1.4), this `continuum softening' must be a common occurrence at higher redshift. This is
indeed suggested in
a recent survey of 49 Seyfert 2 X-ray spectra (Guainazzi \et 2005) which found highly obscured AGN invariably exhibit
a prominent soft excess above the extrapolated hard X-ray power law. Conversely, we conclude the extremely hard 2-10 keV spectrum of \mkn\
shows that such sources cannot be very common, a conclusion consistent with the interpretation in P05 of a Seyfert being viewed
(just) through the edge of the obscuring torus. 

Line fluxes from both neutral and ionised matter yield equivalent widths that
support the continuum deconvolution. However, comparison of the reflected continua with an estimate of the intrinsic nuclear
luminosities shows that both cold and warm reflection is stronger in \mkn, particularly so for the former. The large cold
reflection and marginally Compton thick absorber in \mkn\ led P05 to suggest that \mkn\ was being viewed through the near-edge
of the putative torus. We now find the much lower cold reflection and higher opacity to the nucleus suggests that \ngc\ is being
viewed near  side-on, i.e. close to the plane of the torus. The further outcome of our comparison of the X-ray spectra of \ngc\ and
\mkn, that the total warm reflection is somewhat stronger in \mkn, is also consistent with that explanation, whereby more of the 
outflowing ionised gas can be seen over the edge of the obscuring cold matter. One
caveat on that conclusion, which assumes \ngc\ and \mkn\ are intrinsically similar objects, is raised by the relatively narrow
permitted optical  lines (Miller \et\ 1991) and steep X-ray power law in \ngc, suggesting it harbours a NLS1 nucleus, where a
significantly higher accretion ratio (Pounds \et\ 1995, Laor 1997) could affect the density and distribution of circumnuclear 
matter.   

Finally we note how our analysis may be relevant to the strong soft excesses widely `observed', but physically difficult to 
explain (e.g. Gierlinski and Done 2004) in Type 1 Seyferts. First, taking our interpretation of the stronger warm reflector in
\mkn\ as being due to seeing deeper into the throat of the torus, observing the ${\it total}$ outflow in a type 1 Seyfert might 
provide
the enhanced  soft X-ray emission component discussed in Pounds \et\ (2005). Second, the absence of a significant `upturn' below
$\sim$1 keV in the warm reflected continuum of \ngc\ contrasts with the strong `soft excess' often claimed in Seyfert 1 spectra.
Both these findings support recent suggestions (e.g. Gierlinski and Done 2004, Pounds \et\ 2004a,b) that the enigmatic `soft
excess', a long-standing property of AGN (Turner and Pounds 1989) needs a fundamental re-assessment.

\section{ Acknowledgements } 

The results reported here are based on observations obtained with \xmm, an ESA science mission with instruments and
contributions directly funded by ESA Member States and the USA (NASA). The authors wish to thank the SOC and SSC teams for
organising the \xmm\ observations and initial data reduction, and the referee for a detailed report which improved the focus of
the paper. 
KAP gratefully acknowledges the support of a Leverhulme Trust
Emeritus Fellowship. SAV is supported on a PPARC research grant.

\appendix

\section{Bayesian analysis}
The excess emission at $\sim$6--8 keV shown in the continuum fit to the \ngc\ EPIC spectrum (figure 2) has been resolved into more and
more components as better data have become available. A BBXRT observation (Marshall \et\ 1993) suggested 3 components, identified
with Fe K$\alpha$, He$\alpha$ and Ly$\alpha$, a view supported by an \asca\ observation from which Iwasawa \et\ (1997) added
a fourth component identified with the Compton shoulder of the Fe K$\alpha$ line. Making first use of the present \xmm\ data Matt04 
analysed the pn spectrum between 4--10 keV and identified a total of 9 lines, 5 from cold and 4 from warm matter.Our
analysis resolved a similarly complex spectrum and we examine the statistical basis for such spectral line fits in this Appendix.
   
Justifying the number of lines included in a spectral fit is a challenging problem for classical (i.e. frequentist) statistics.
Given  an X-ray spectrum comprising a continuum and a ``few'' emission lines, how many
lines are there and what are their energies and strengths?  Fortunately there is a Bayesian solution to this problem
that gives the relative probabilities of there being $M$ lines in a given spectrum.  The method was
discussed in detail by Sivia \& Carlile (1992; see also section 4.2 of Sivia 1996), where it was applied to molecular
spectroscopy data. We briefly review the method before presenting its results.

We require the posterior probability, denoted $p(M | \mathbf{D},I)$, of there being $M$ lines in a
spectrum based on an analysis of the data array $\mathbf{D}$,  where $I$ symbolises all the background information necessary for
the analysis (i.e., that the data have Gaussian errors, how to specify the continuum model, and that the instrument calibration
is well understood). In order to calculate $p(M | \mathbf{D}, I)$ we employ Bayes theorem to ``invert'' the integrated
likelihood $p(\mathbf{D} | M, I)$, which itself can be derived using the results of standard maximum likelihood (ML) analysis
(in this case $\chi^2$-minimisation).

The procedure works by adding lines to the model one at a time, obtaining the ML solution each time, and recording the number of
parameters, the $\chi^2$ and covariance matrix ${\mbox{\boldmath $\sigma$}}^2$ of the fit. Initially the $M=0$ model is fitted, 
i.e. the underlying continuum model (in this case an absorbed power law
and PEXRAV cold reflection continuum). Then a Gaussian of fixed width is added to the model  and the change
in $\chi^2$ is measured as the line energy is stepped through the range $6-10$~keV in $0.1$~keV intervals, allowing the
normalisation to vary at each position. The line energy giving the minimum $\chi^2$ is
taken to be a good initial guess for the ML line energy, and the model is then fitted, starting from this position, to find the
best $\chi^2$. The process is repeated, at each successive iteration adding one more line.
For each value of $M$ the following expression is evaluated:
\begin{equation}
p(M | \mathbf{D}, I) \propto
\frac{ M! (2\pi)^{P/2} \mathcal{L}_0 \sqrt{\det[{\mbox{\boldmath $\sigma$}}^2]}
}{ 
([ E_{\rm max} - E_{\rm min}] A_{\rm max})^M }
\label{eqn:bayes-covar}
\end{equation}

The term $([ E_{\rm max} - E_{\rm min}] A_{\rm max})^M$ represents the volume of parameter space searched in order to find $M$
lines, where $E_{\rm min}$ and $E_{\rm max}$ are the lower and upper energy bounds placed on the lines ($6$ and $10$~keV,
respectively) and $A_{\rm max} = 10^{-4}$~ph cm$^{-2}$ s$^{-1}$ is the maximum normalisation allowed in the fit. Together these
prescribe our prior ignorance about the line energies and normalisations, i.e. prior to examining the data we make the most 
least-informed assumption and consider lines to be equally probable anywhere in the range $E_{\rm min} - E_{\rm max}$ with any
normalisation $A < A_{\rm max}$.  In the calculation of $p(M | \mathbf{D}, I)$ this term effectively acts as a penalty for 
adding extra parameters  in the model, enacting ``Occam's razor.''

The term $(2\pi)^{P/2} \mathcal{L}_0\sqrt{\det[{\mbox{\boldmath $\sigma$}}^2]}$  gives the integrated likelihood of the data
given a model containing $M$ lines and $P$ free parameters (in our case $P=3+2M$ since the continuum has three free parameters
and each Gaussian line has two). The likelihood value at the best fit is $\mathcal{L}_0 \propto \exp(-\chi^2/2)$. The implicit
assumption is that the likelihood surface can be approximated by a $P$-dimensional Gaussian, and this expression is then the
integral of that Gaussian with a peak of $\mathcal{L}_0$ and a shape (widths) defined by the covariance matrix ${\mbox{\boldmath
$\sigma$}}^2$, calculated using the Laplace approximation\footnote{The validity of this assumption can be checked by comparing
the errors on each parameter calculated from the diagonal elements of the covariance matrix, $\sqrt{\sigma_{xx}^2}$, with the
errors calculated using a $\Delta \chi^2= 1.0$ criterion. If the two error estimates are in agreement for each parameter then
the likelihood surface is Gaussian and the Laplace approximation is valid.}.

Finally, the factor of $M!$ simply accounts for the degeneracy in the fit: the order of the $M$ lines is irrelevant and so there
are $M!$ identical maxima in the likelihood space corresponding to all the possible permutations.

Once we have computed $p(M | \mathbf{D}, I)$ for $M=0,1,2,\ldots,M_{\rm max}$ we can then normalise them such that $\sum_{M} p(M
| \mathbf{D}, I) = 1$.  In doing so we have implicitly  assigned an equal prior probability to the each value of $M$, in other
words we admit total ignorance about the number of lines in the spectrum  except to define an upper limit (in this case $M \le
13$).  No value of $M$ is any more probable than any other. This factor, $1/(1+M_{\rm max})$, is present in the constant of
proportionality above but can be ignored since the results are normalised.

\begin{figure}
\begin{center}
\hbox{\includegraphics[width=6 cm, angle=270]{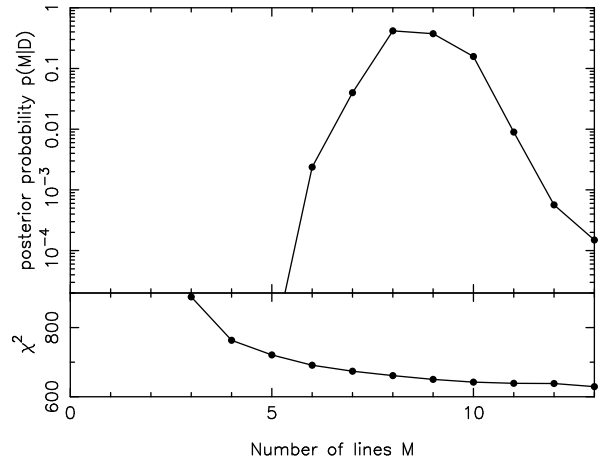}}
\end{center}
\caption{Posterior probabilities for their being $M$ lines in the $6-10$~keV EPIC pn spectrum of NGC 1068. The probabilities were
calculated for $M=0,1,2,\ldots,13$ lines and re-normalised such that $\sum_{M} p(M | \mathbf{D}, I) = 1$. The bottom panel shows
the corresponding $\chi_{\rm min}^2$ for each model.}
\end{figure}

Using Xspec to fit in sequence $M=0,1,2,\ldots,13$ lines to the data, including a
power law plus cold reflection continuum, the above method was applied to the $3.5-15$~keV EPIC pn spectrum of \ngc. For each model 
the posterior probability was calculated, and the results
are shown in figure 8. The most probable number of lines is $8$ but there is little to favour this over $9$ or $10$
lines. However, $M=8-10$ lines are quite strongly preferred over fewer lines,
with odds of $p(M=8-10)/p(M<8) = 22.5:1$. There is also no evidence to favour more lines, with odds of $p(M>10)/p(M=8-10) =
1:98$ against there being more than $10$ lines.

This method therefore provides reliable evidence for at least $8$ and as many as $10$ separate line-like emission components in
the EPIC spectrum of NGC 1068, and is free from the statistical problems often associated with detecting emission lines (see
discussion in Protassov et al. 2002). In
passing we note that Freeman et al. (1999) previously used essentially the same method (based on identical logic)  to assess the
evidence for absorption lines in the X-ray spectra of a $\gamma$-ray burst.

\bsp 
\label{lastpage}

\end{document}